\documentclass[conference,compsoc]{IEEEtran}

\usepackage{cite}
\usepackage{flushend}
\usepackage{todonotes}
\usepackage{amsmath}

\begin{document}
\title{Cost Explosion for Efficient Reinforcement Learning Optimisation\\of Quantum Circuits}

\author{
\IEEEauthorblockN{Ioana Moflic}
\IEEEauthorblockA{Aalto University, Espoo, Finland\\
ioana.moflic@aalto.fi}
\and
\IEEEauthorblockN{Alexandru Paler}
\IEEEauthorblockA{Aalto University, Espoo, Finland\\
alexandru.paler@aalto.fi}
}

\maketitle

\begin{abstract}
Large scale optimisation of quantum circuits is a computationally challenging problem. Reinforcement Learning (RL) is a recent approach for learning strategies to optimise quantum circuits by increasing the reward of an optimisation agent. The reward is a function of the quantum circuit costs, such as gate and qubit counts, or circuit depth. Our goal is to improve the agent's optimization strategy, by including hints about how quantum circuits are optimized manually: there are situations when the cost of a circuit should be allowed to temporary explode, before applying optimisations which significantly reduce the circuit's cost. We bring numerical evidence, using Bernstein-Vazirani circuits, to support the advantage of this strategy. Our results are preliminary, and show that allowing cost explosions offers significant advantages for RL training, such as reaching optimum circuits. Cost explosion strategies have the potential to be an essential tool for RL of large-scale quantum circuit optimisation.
\end{abstract}

\IEEEpeerreviewmaketitle

\section{Introduction}

Large scale quantum circuit compilation and optimisation are a necessity for achieving the goal of practical quantum computations. There is a gap between the capabilities of the state of the art quantum circuit optimisation tools and the size of practical circuits. Current compilers can handle circuits with tens of qubits and hundreds of gates, while practical  circuits operate on thousands of qubits and include orders of magnitude more gates.

Machine learning techniques are applied to quantum circuit compilation and optimisation (e.g.~\cite{paler2020machine, zulehner2019evaluating}). The general approach is to invest large amounts of computational power into the training of models that can then be used for fast and efficient quantum circuit compilation. Reinforcement Learning (RL) is one of the machine learning approaches that have been proposed.

\begin{figure}[!t]
    \centering
    \includegraphics[width=0.99\columnwidth]{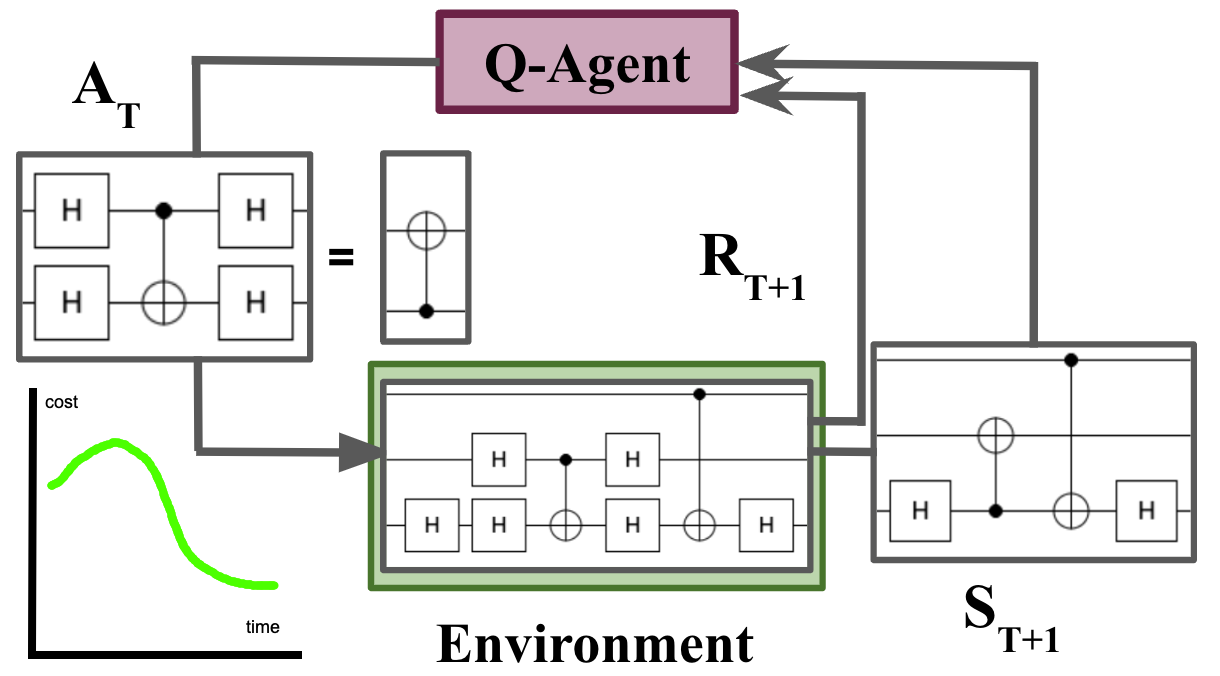}
    \caption{The learning framework of the RL framework employs an agent whose goal is to interact with the environment (the quantum circuit) by choosing actions that maximize its rewards. At a given time step $T$, an agent chooses an action $A_T$ and receives reward $R_{T+1}$ and the next state $S_{T+1}$ of the environment.}
    \label{fig:arch}
\end{figure}

RL is a method in which an agent uses a trial and error approach in an iterative manner with the goal of finding the optimal policy that solves a specific problem~\cite{Sutton1998}. The general approach is to give the agent full information about the environment, which is the quantum circuit, and let it interact with the environment by applying various circuit gate transformations until it manages to find a sequence that reduces the overall cost of the circuit (Fig.~\ref{fig:arch}).

\begin{figure*}[!t]
\centering
\includegraphics[width=\textwidth]{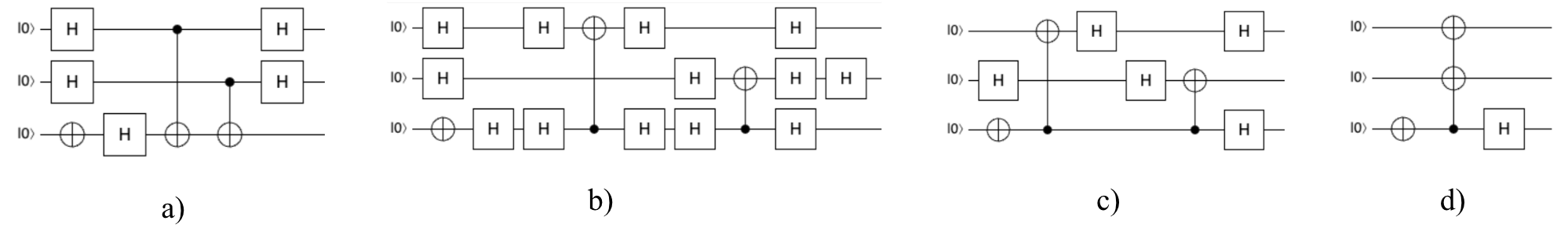}
  \caption{The manual optimization procedure for a three-qubit Bernstein-Vazirani quantum circuit requires the agent to flip all CNOTs. The initial circuit in a) has a lower depth than the one in b), where all CNOT gates were flipped. In c), Hadamard gates are cancelled to allow for the parallelization of the CNOT gates that yields the optimal form of the circuit in d). However, during the training procedure, the agent will need to perform a single CNOT flip or cancellation at a given time step, which will result in a total of ten circuit transformations applied successively.}
  \label{fig:motiv}
\end{figure*}

\subsection{Background}
\label{sec:met}

This section presents the details necessary for describing our RL framework and approach. Our RL tool is using the Q-Learning algorithm to train an agent on how to apply actions in the form of circuit template rewrite rules (Fig.~\ref{fig:motiv}).

Q-Learning is a simple, but effective RL algorithm used to solve Markov Decision Processes (MDP). The agent starts with zero information about the MDP, but it learns the transition probabilities by choosing actions and learning the associated states and the corresponding rewards. The goal of the agent is to increase its long term rewards and it does so by learning the values $Q(s,a)$ (or q-values), which give the "value" of taking the action $a$ from state $s$. By the end of the training, it is assumed that q-values converge to the expected value of taking action $a$ from state $s$. The values $Q(s,a)$ are stored in a data structure known as Q-Table. The size of the Q-Table increases faster at the beginning of learning, when the agent is \emph{exploring} the environment. The Q-Table size increase is slower towards the end of the learning: the agent will \emph{exploit} the knowledge it accumulated.

At each training step, a reward value is computed as a response from the environment. The value of the reward depends on the action chosen by the agent (\emph{i.e.} which circuit transformation is being selected).

Quantum circuit optimisation can be tackled by using Reinforcement Learning. The circuit is a fully observable environment and gate identities that preserve logical equivalence are actions an agent can choose from when learning an optimisation policy. The action and state spaces are discrete, but size-dependent on the structural constraints of the circuit. The agent will use template rewrite rules as actions.

\begin{figure}[!h]
    \centering
    \includegraphics[width=0.75\columnwidth]{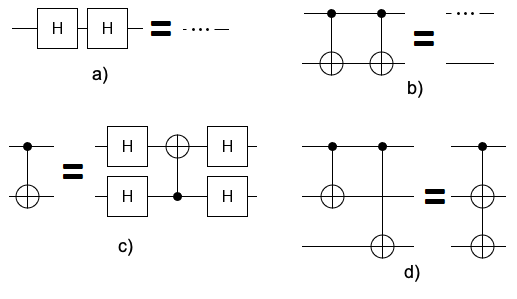}
    \caption{Quantum circuit template-based rewrite rules. a) two neighbouring Hadamard gates are canceled; b) two neighbouring CNOT gates are canceled; c) a CNOT surrounded by two Hadamard gates is reversing the direction of the CNOT; d) CNOTs sharing the same control qubit can be parallelized.}
    \label{fig:rules}
\end{figure}

Template rewrite rules (e.g. Fig.~\ref{fig:rules}) are a set of circuit equivalences that, if combined efficiently, can optimise quantum circuit depth by allowing gate cancellation or parallelisation to occur. Both gate cancellation and parallelism reduce the error probability Gate cancellations reduce the error probability of executing the circuit. Gate parallelism reduces the run time of the circuit. The latter is especially useful for Noisy Intermediate Scale Quantum (NISQ) as well as next-generation error-corrected machines.

The cost of a quantum circuit does not have a straightforward definition. Usually it refers to the number of gates~\cite{miller2003transformation, saeedi2013synthesis}, but it can also include the depth of the circuit. At times, when some gates are considered more \emph{expensive} than others, each gate has its own cost and the total cost is the sum of the individual costs. Expensive gates are T gates in the context of Clifford+T quantum circuits which are compiled for error-correction~\cite{paler2022realistic}. For the case of NISQ circuits, expensive gates might be the SWAP gate, which is necessary for implementing long-range interactions (e.g.~\cite{zulehner2018efficient}).

\begin{figure*}[!t]
    \centering
    \includegraphics[width=0.45\textwidth]{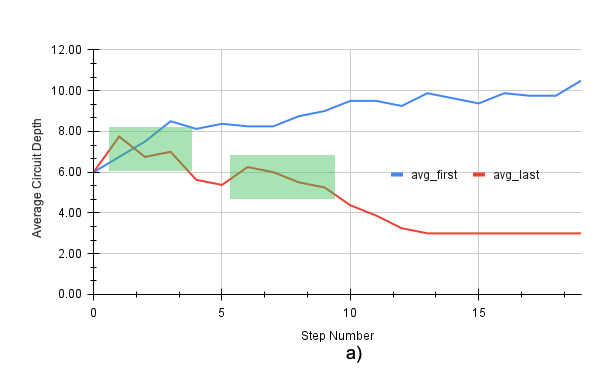}
    \includegraphics[width=0.41\textwidth]{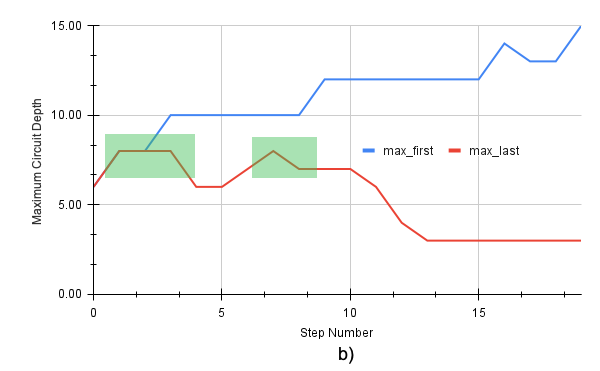}
    \caption{Circuit depths during the execution of RL training on a Bernstein-Vazirani circuit operating on three qubits. The optimization is learned during a sequence of episodes by an agent. We plot the circuit depth observed in the first episode (blue line, when the agent has no experience) and the last episode (red, when the agent has learned the optimization heuristic). a) the average of the depth values observed per episode step; b) the maximum of the depth values observed per episode step. The green regions indicate where TCE takes place: two times during the optimization of this circuit.}
    \label{fig:bv3}
\end{figure*}

\subsection{Motivation}

Existing RL frameworks for quantum circuit optimisation cannot handle large-scale circuits~\cite{fosel2021quantum, li2023quarl, van2023qgym}. On the one hand, this is because RL is a very computationally intensive method and training it on large circuits requires very large amounts of compute power (e.g. supercomputers~\cite{li2023quarl}). On the other hand, RL is starting from zero knowledge about the problem being solved and discovering new heuristics by the agent consumes a lot of time spent with trial-and-error.

A seemingly breakthrough result published by Google~\cite{mirhoseini2021graph} shows that it is possible to scale RL for realistic classical circuit sizes. However, the result is under scrutiny~\cite{markov2023false}. Nevertheless, from a diagrammatic perspective (quantum circuits represented as directed acyclic graphs), classical circuits share similarities with quantum ones. Assuming that Google's result is valid, there are no reasons to expect that RL optimization of quantum circuits would not scale.

Moreover, for the particular case of RL for quantum circuit optimisation, the generally accepted heuristic of continuously trying to lower a cost function does not resemble the way circuits are optimised in practice. Sometimes, in order to achieve optimal circuits, the cost should be allowed to increase in order to have the circuit in a form that allows the application of an optimization sequence which eventually lowers the cost. 

We call \emph{Temporary Cost Explosion} (TCE) the optimisation strategy of allowing the cost to increase for a short period of time during training. The observation on which TCE is built has recently appeared as a side note in~\cite{li2023quarl}. However, to the best of our knowledge, our work is the first in which TCE is described, analysed and benchmarked. We present an example of TCE in Fig.~\ref{fig:motiv}, where an agent is repeatedly flipping CNOT gates by first adding four adjacent single qubit gates, and then cancelling neighbouring ones.

\subsection{Contribution}

We focus on engineering the reward function in order to improve RL (for quantum circuit optimization). We have chosen to control the agent's behaviour through the reward function, instead of focusing on policies that encourage novelty~\cite{conti2018improving}, or on organizing the optimization tasks of the agent according to a curricula\cite{klink2021probabilistic}. Our choice of implementation is based on its simplicity while delivering very encouraging results in practice. We employ Q-Learning as it uses fewer hyperparameters compared to deep RL. This enables a clearer analysis of the importance of the reward function with respect to number of environment states~\cite{moflic2023graph} as well as convergence behaviour. 

To the best of our knowledge, this is the first work which analyzes the influence of reward cost explosions on the efficiency of RL for quantum circuit optimization. For example, in the context of VLSI classical circuits, similar strategies have been used for place and route by simulated annealing~\cite{markov2023false, sechen1985timberwolf}. This work extends the state-of-the-art regarding scalable quantum circuit compilation methods in the following ways:
\begin{itemize}
    \item proposes to allow for TCE in order to achieve lower cost circuits;
    \item presents an example of how to encode TCE into reward functions;
    \item implements the novel reward function and numerically demonstrates its utility.
\end{itemize}

\begin{figure*}[!t]
    \centering
    \includegraphics[width=0.45\textwidth]{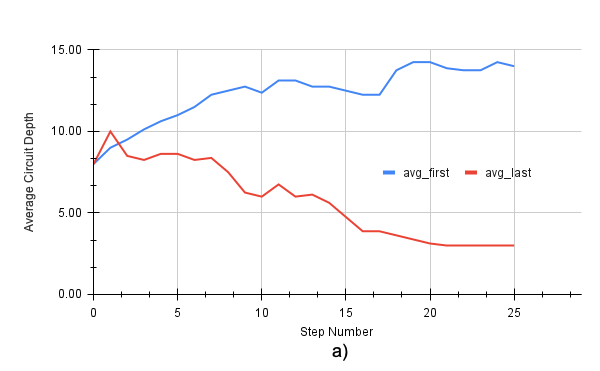}
    \includegraphics[width=0.41\textwidth]{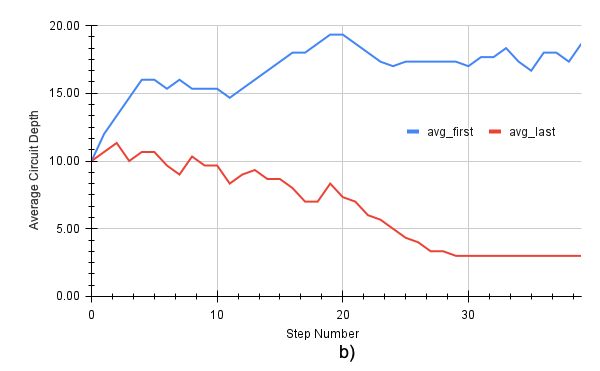}
    \caption{Average circuit depths during the execution of RL training on Bernstein-Vazirani circuits operating on five and seven qubits. These plots are similar to Fig.~\ref{fig:bv3}a. When counting the peaks of the red line, we observe that: a) for the five qubit circuit, TCE is applied on average automatically for three times, b) for the seven qubit, TCE is applied for five times.}
    \label{fig:bv57}
\end{figure*}

\section{Methods}

RL for quantum circuit optimisation is equivalent to learning a heuristic for applying template rewrite rules. The RL procedure is focused on an agent that is observing an environment (the quantum circuit to be optimised). Learning is performed in episodes, and each episode is composed of multiple steps. The number of episodes and the number of steps are hyper-parameters which influence the convergence of the learning procedure (Section~\ref{sec:res}).

Each step of an episode is equivalent to applying one action (rewrite rule) on the circuit and retrieving the corresponding reward. The reward is computed after executing a function that analyses the output circuit and returns a list of costs. For example, assuming that a step is the application of the rewrite rules that take the circuit from Fig.~\ref{fig:motiv}a) to Fig.~\ref{fig:motiv}b), the costs of the latter circuit will be, for example, \emph{depth=9}, \emph{gate-count=15}. This is in contrast to the input circuit, where \emph{depth=5}, \emph{gate-count=8}. There is a \emph{cost explosion} taking place as a result of this step.

It seems contradictory to allow cost explosions while increasing cumulative rewards. During an episode, the goal of the agent is to maximize the cumulative reward. Usually, the reward is an inverse relation to a cost. For example, higher depth implies a low reward, and lower depth generates a higher reward. However, this is possible, and we will sketch a solution in the following.

Our goal is to steer the behaviour of the agent through the reward function. While local optimizations could be applied for this type of circuit, we believe that large-scale circuit optimisation requires the agent to learn an algorithm rather than local optimisation patterns. We need to design a reward function that allows for TCE to occur and also reach the optimal form of the circuit. In order for this to happen, the reward function must be able to encode the optimisation algorithm, which the agent will later decode in order to find the optimal policy. We propose an exponential reward function of the form:

\begin{align*}
reward_T(C) = \left(\prod_i^n \frac{k_i}{cost_i(C)}\right)^{\left(\sum_j^m \frac{k_j}{cost_j(C)}\right)}
\end{align*}

In the previous expression, the constants $n$ and $m$ are the number of optimisation criteria assigned to the base and exponent. The values $cost_i$ and $cost_j$ stand for two different costs recorded at a given time-step $T$ for circuit $C$. For example, a simple version of the reward function can have $n=1, m=1$. In this situation, both base and exponent refer to a single cost type, such as gate count ($cost_i$) and circuit depth ($cost_j$). Parameters $k_i$, $k_j$ control the ratios in the basis and exponent.

\begin{figure}[!t]
    \centering
    \includegraphics[width=0.7\columnwidth]{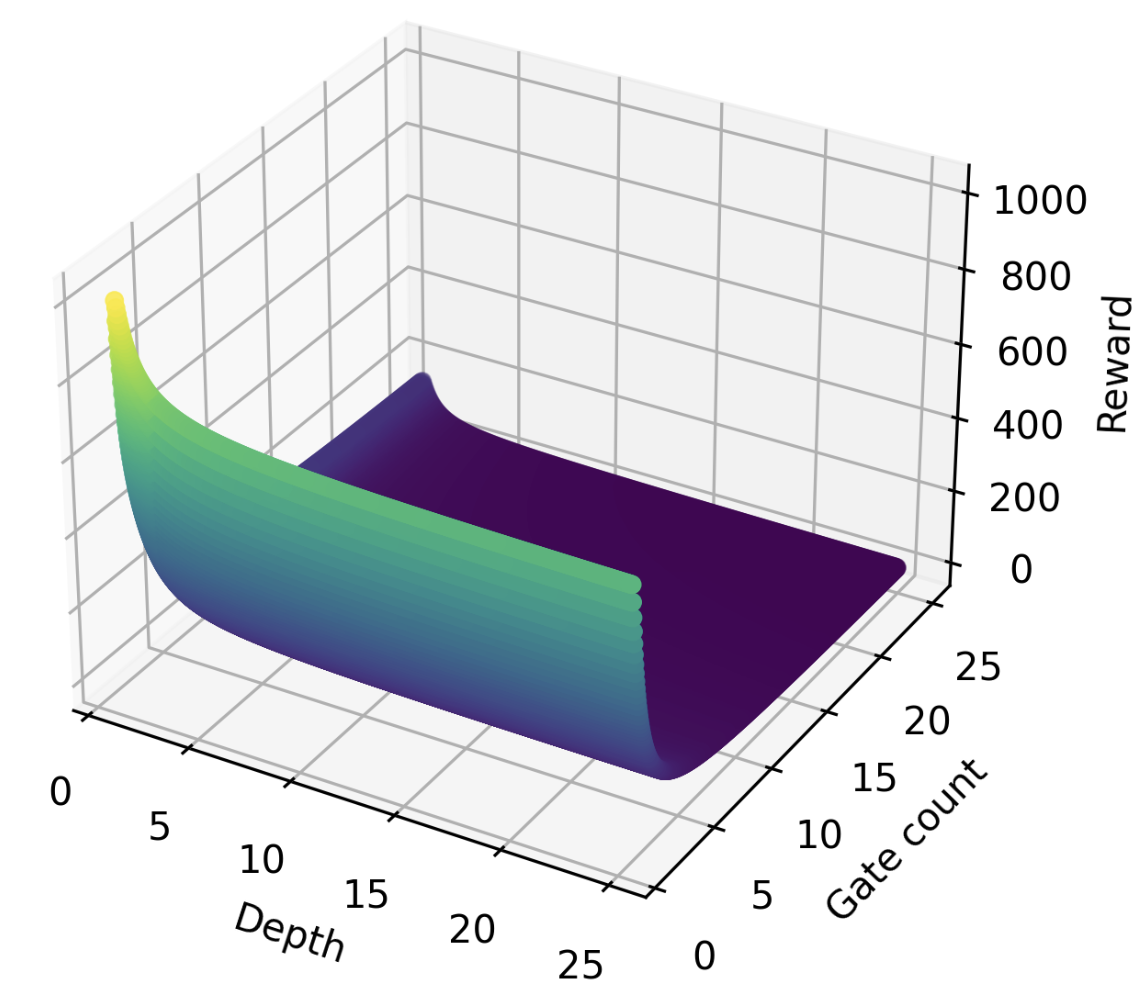}
    \caption{Example of an exponential reward function. This function focuses on the optimisation of gate count and depth. The reward peaks when both the depth and the gate count are low.}
    \label{fig:reward3D}
\end{figure}

Fig.~\ref{fig:reward3D} illustrates how such an exponential function behaves for gate count and depth. For the simplified episode example from Fig.~\ref{fig:motiv}, by applying our exponential reward function, we observe the values from the table below.

\begin{table}[!h]
    \centering
    \begin{tabular}{c|c|c|c}
    Circuit & Gates & Depth	& Reward\\
     \hline
    1       & 8     & 5	    & 0.8706\\
    2       & 16    & 9     & 0.8572\\
    3       & 8     & 6     & 0.8909\\
    4       & 3     & 3     & 1.1006
    \end{tabular}
\end{table}

Our reward function is different from other reward functions in the RL literature for quantum circuit optimization, where the rewards are either subtractions or divisions of cost values resulted from two consecutive steps. Such functions do not reward agents which repeatedly apply optimisation template rules that increase the gate count of the circuit -- TCE cannot be performed.

Our exponential reward function makes a distinction between the costs in the base and the costs in the exponent. The increase of both $cost_i$ and $cost_j$ lowers the reward, but there will be more emphasis put on reducing the cost at the exponent. This translates to the idea that for example, adding gates does not have a great effect on the reward as long as the depth is not increased. For the circuit in~\ref{fig:motiv}b) the increase in the number of gates could increase the cost, but this increase will be compensated by the decreased depth -- however, this compensation depends on the value of $k$.

\section{Results}
\label{sec:res}

We illustrate the potential of TCE by benchmarking it with Bernstein-Vazirani circuits. These circuits have a known optimal depth of three. We use Q-Learning to learn the optimisation of the circuits. The Bernstein-Vazirani circuits are particularly challenging to optimize because every CNOT in the circuit must be flipped in order for future Hadamard gate pairs to be cancelled. Even though there are multiple optimal policies that the agent can follow, each requires the depth of the circuit to first increase and then decrease in later stages of the optimisation procedure.

Our empirical results are obtained by: 1) repeating the training for 10 times, 2) collecting in each run the depths at each step of the first and last episode, and 3) then calculating the average depth per step. During training we used $n=1, m=1$, where $cost_i$ is a function of gate count and $cost_j$ is circuit depth. The gate cost is computed as a weighted sum of the gates in the circuit. It is assumed that multi-qubit gates are more expensive than single qubit gates. The values of $k_1$ and $k_2$ used in the reward function are tuned per epoch to account for the newly gained experience of the agent.

\begin{figure}[!t]
    \centering
    \includegraphics[width=0.75\columnwidth]{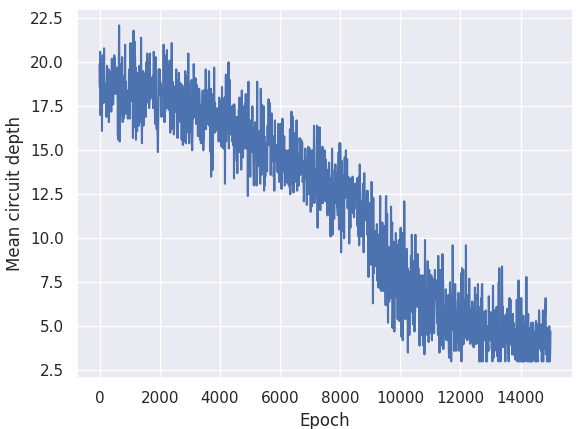}
    \caption{The depth of the seven qubit Bernstein-Vazirani circuit reaches optimum after 10000 episodes with the given reward function.}
    \label{fig:depth3}
\end{figure}

\begin{figure}[!t]
    \centering
    \includegraphics[width=0.75\columnwidth]{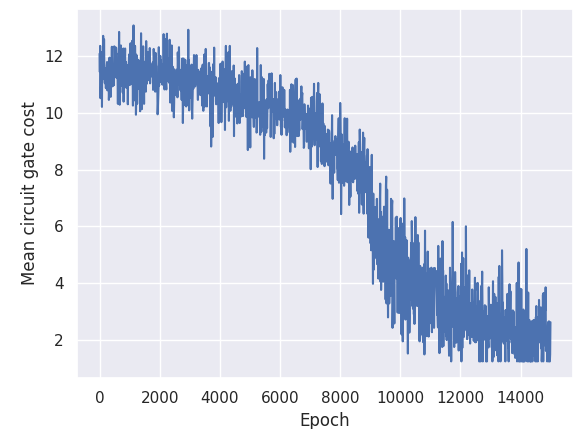}
    \caption{The gate cost of the seven qubit Bernstein-Vazirani circuit reaches optimum after 10000 episodes with the given reward function.}
    \label{fig:cost3}
\end{figure}

Fig.~\ref{fig:depth3} illustrates an example of the circuit depth during the entire training procedure. We observe that in the presence of TCE, the depth is reduced until it reaches its optimal value. This shows that the cumulative reward per episode is maximized. However, for validation purposes, we are interested in tracking the evolution of the circuit depth within an episode, too. Fig.~\ref{fig:cost3} shows a similar trend to Fig.~\ref{fig:depth3}: the gate cost drops significantly around epoch 10000. The circuit depth plots show that TCE is successfully applied on multiple occasions during the optimisation. The agent seems to prefer to perform optimizations greedily: there is a sequence of small TCE (green areas in Fig.~\ref{fig:bv3}) instead of a single large TCE followed by a sequence of optimisations.

Figs.~\ref{fig:bv3} and~\ref{fig:bv57} illustrate the observed depth reduction during the RL training of an agent on Bernstein-Vazirani circuits of 3, 5 and 7 qubits. Table~\ref{tab:config} summarizes the hyper-parameters used for training the optimisation agent, where $LR$ is the learning rate and the \emph{exploration decay} is selectively decreased with the size of the circuit.

\begin{table}[!h]
    \centering
    \caption{Q-Learning Hyper-parameters}
    \label{tab:config}
    \begin{tabular}{c|c|c|c|c|c}
    Circuit& LR & Episodes & Expl. Decay & Disc. factor & Steps\\
    \hline
    BVZ-3   & 1e-3  & 3000  &   1e-3 &  0.97-0.98    &30\\
    BVZ-5   & 1e-3  & 3000  &   9e-4 &  0.97-0.98    &35\\
    BVZ-7   & 3e-3  & 15000 &   9e-5 &  0.97-0.98    &50\\
    \end{tabular}
\end{table}

\section{Conclusion}

We presented TCE, which is a way of optimising circuits by temporarily allowing the circuits to have worse costs. In order to implement TCE we developed an exponential reward function. We used Bernstein-Vazirani circuits to illustrate the utility of TCE. Future work will focus on training RL on larger and more diverse benchmark circuits, and on comparing the convergence times of our approach with more complex methods such as~\cite{conti2018improving, klink2021probabilistic}.

\section*{Acknowledgment}
This research was developed in part with funding from the Defense Advanced Research Projects Agency [under the Quantum Benchmarking (QB) program under award no. HR00112230007 and HR001121S0026 contracts]. We would like to thank Niki Loppi of the NVIDIA AI Technology Center Finland for his help with the implementation. We acknowledge the funding received from the Finnish-American Research and Innovation Accelerator, one of eight global pilots funded by the Finnish Ministry of Education and Culture.

\bibliographystyle{IEEEtran}
\bibliography{__main}

\end{document}